\documentclass[conference]{IEEEtran}
\IEEEoverridecommandlockouts
\usepackage{cite}
\usepackage{amsmath,amssymb,amsfonts}
\usepackage{algorithmic}
\usepackage{graphicx}
\usepackage{textcomp}
\usepackage{xcolor}
\usepackage{comment}
\usepackage{breakurl}
\usepackage{xspace}
\usepackage{pifont}
\usepackage{multirow}
\usepackage{amsfonts}
\usepackage{listings}
\usepackage{xcolor}
\usepackage{color}
\usepackage{url}
\usepackage{cite}
\usepackage{booktabs}

\usepackage{hyperref}
\newcommand{\RQone}{ Can our proposed  method {\tool} outperform the state-of-the-art baselines in the task of Bash code comment generation?}
\newcommand{\RQtwo}{Can two-stage training strategy help for improving the performance of our proposed method {\tool}?}
\newcommand{\RQthree}{How effective are the component settings of our proposed method {\tool}?}

\definecolor{dkgreen}{rgb}{0,0.6,0}
\definecolor{gray}{rgb}{0.5,0.5,0.5}
\definecolor{mauve}{rgb}{0.58,0,0.82}

\newcommand{\tool}{\textsc{BashExplainer}\xspace}
\newcommand{\stageone}{Bash code encoder construction stage}
\newcommand{\stagetwo}{Bash code comment generation stage}

\newcommand{\yc}[1]{\textcolor{black}{#1}}
\newcommand{\modify}[1]{\textcolor{black}{#1}}

\begin{document}

\title{{\tool}: Retrieval-Augmented Bash Code Comment Generation based on Fine-tuned CodeBERT}

\author{\IEEEauthorblockN{Chi Yu\IEEEauthorrefmark{2},  
Guang Yang\IEEEauthorrefmark{2}, 
Xiang Chen\IEEEauthorrefmark{2}\IEEEauthorrefmark{1},  
Ke Liu\IEEEauthorrefmark{2}, Yanlin Zhou\IEEEauthorrefmark{2}}
\IEEEauthorblockA{\IEEEauthorrefmark{2}\textit{School of Information Science and Technology},
\textit{Nantong University}, China\\
Email: yc\_struggle@163.com, novelyg@outlook.com, xchencs@ntu.edu.cn, aurora.ke.liu@outlook.com, 1159615215@qq.com}
}

\maketitle

\begingroup
\renewcommand{\thefootnote}{}
\footnotetext[1]{\IEEEauthorrefmark{1} Xiang Chen is the corresponding author.}
\endgroup

\begin{abstract}

Developers use shell commands for many tasks, such as file system management, network control, and process management. Bash is one of the most commonly used shells and plays an important role in Linux system development and maintenance. Due to the language flexibility of Bash code, developers who are not familiar with Bash often have difficulty understanding the purpose and functionality of Bash code.
In this study, we study Bash code comment generation problem and proposed an automatic method {\tool} based on two-stage training strategy.
In the first stage, we train a Bash encoder by fine-tuning CodeBERT on our constructed Bash code corpus. In the second stage, we first retrieve the most similar code from the code repository for the target code based on semantic and lexical similarity. Then we use the trained Bash encoder to generate two vector representations. Finally, we fuse these two vector representations via the fusion layer and generate the code comment through the decoder. To show the competitiveness of our proposed method, we construct a high-quality corpus by combining the corpus shared in the 
previous NL2Bash study and the corpus shared in the NLC2CMD competition. This corpus contains 10,592 Bash codes and corresponding comments. Then we selected ten baselines from previous studies on automatic code comment generation, which cover information retrieval methods, deep learning methods, and hybrid methods. The experimental results show that in terms of the performance measures BLEU-3/4, METEOR, and ROUGR-L, {\tool} can outperform all baselines by at least 8.75\%, 9.29\%, 4.77\% and 3.86\%. Then we design ablation experiments to show the component setting rationality of {\tool}. Later, we conduct a human study to further show the competitiveness of {\tool}. Finally, we develop a browser plug-in based on {\tool} to facilitate the understanding of the Bash code for developers.
\end{abstract}


\section{Introduction}

Shell is the interface between the developer and the Linux operating system.  Nowadays, Linux operating system supports different types of shells.
Bash is the default shell command language for Linux and has been widely used in Linux development and maintenance.
Compared with the advanced programming languages (such as  Java, Python), Bash language is used in a few scenarios, but the role of the Bash language during the Linux system development and maintenance still cannot be ignored.

As a script language, the Bash language has the feature of language flexibility~\cite{lin2018nl2bash}. Therefore, for developers who are not familiar with the Bash language, it is still challenging for them to understand Bash code during the system development and maintenance. Based on our statistical analysis,
there are 86,846 Q\&A posts related to the keyword ``shell" and 143,743 Q\&A posts related to the keyword ``Bash" on Stack Overflow until March 2022. 
In Table~\ref{tab:examples}, we use a post\footnote{https://stackoverflow.com/questions/56229939/what-does-set-2-mean-in-bash-shell} from Stack Overflow to show the necessity of automatically generating comments for Bash codes.
According to the content of this post, we find that this user could not understand the meaning of this Bash code. 
Therefore, it is challenging for developers who are not familiar with the Bash language to understand the purpose and functionality of Bash code. 

\begin{table}[]
\caption{A Post Related to Bash Code understanding on Stack Overflow}
 
\begin{tabular}{ll}
\toprule
\textbf{Post Title} &what does `` set -- \$\{@ : 2\} " mean in Bash shell?        \\
\midrule
\textbf{Content} & \begin{tabular}[c]{@{}l@{}}
I have a shell script, \\
and a line in that is ``set -- \$\{@ : 2\}", \\ 
could you please tell me what does it mean? \\
I have tried in my test script, \\
it seems to remove the \$1 param in args, \\
could you help me with more details? 
\end{tabular}\\
\midrule
\textbf{Tags}                                      & Bash, shell           \\
 \bottomrule
\end{tabular}

\label{tab:examples}
\end{table}

High-quality code comments can improve the readability and comprehensibility of the code and play an important role during software development and maintenance~\cite{sridhara2010towards,wei2020retrieve}. 
However, developers often forget to write comments or do not keep the code comments up to date, which can result in missed or outdated code comments. Previous study~\cite{xia2017measuring} found that understanding source code may take more than half of the time in software development. Therefore, automatic code comment generation is an important research topic in current software engineering research. 
However, to our best knowledge, most of previous studies~\cite{wei2020retrieve,liu2018neural,hu2018deep,yang2021comformer,hu2021automating,li2022setransformer,yang2022ccgir,li2021secnn} mainly focused on source code comment generation for popular programming languages (such as Java and Python), while less attention has been paid to the script language Bash.

In this study, we are the first to study this problem and propose a novel method {\tool} for Bash code comment generation, which uses both an information retrieval method and a deep learning method. Specifically, {\tool} includes two-stage training strategy: {\stageone} and {\stagetwo}. \modify{In the first stage, to better generate code vector representation, we use our constructed corpus to train Bash encoder by fine-tuning CodeBERT~\cite{feng2020codebert}.}
In the second stage, we first utilize an information retrieval module to retrieve the most similar code for the target code. The information retrieval module is designed based on semantic similarity and lexical similarity. Then we input the target code and similar code into the Bash encoder trained in the first stage, which will output two code vector representations. Later, we use normalization operation to process these two vector representations and  fuse these two vectors via the fusion layer. Finally, the fused vectors are passed through the decoder to generate the corresponding code comment.

To evaluate the effectiveness of our proposed method {\tool}, we constructed a high-quality corpus based on the corpus shared in the previous study NL2Bash~\cite{lin2018nl2bash} and the official data from the NLC2CMD competition\footnote{\url{https://eval.ai/web/challenges/challenge-page/674/leaderboard/1831}}. We merged these two corpora and removed duplicate data from them. Notice we only focused on the 135 most useful utilities identified by Linux users by following the study of NL2Bash~\cite{lin2018nl2bash}.

Since we are the first to study the problem of automatic Bash code comment generation, we select the state-of-the-art methods from the previous studies on automatic code comment generation as the baselines for {\tool}. These baselines cover deep learning-based methods~\cite{gu2016incorporating,vaswani2017attention,feng2020codebert,iyer2016summarizing}, information retrieval-based methods~\cite{wei2020retrieve,liu2018neural,haiduc2010use,haiduc2010supporting} and hybrid methods~\cite{hu2020deep,zhang2020retrieval}. The results of the automated evaluation and the human study show that our proposed method {\tool} can achieve better performance than all the baselines. Finally, we also designed ablation experiments to verify the component setting rationality in our proposed method {\tool}.

The main contributions of our study can be summarized as follows.

\begin{itemize}
\item We are the first to study the automated Bash code comment generation problem. 
Then we propose a retrieval-augmented Bash code comment generation method {\tool} based on fine-tuned CodeBERT.

\item We constructed and shared a high-quality corpus for Bash code comment generation. Then we conducted extensive experiments on this corpus to evaluate the effectiveness and the component setting rationality of our proposed method. Both automated evaluation and human study show that {\tool} can outperform state-of-the-art baselines.

\item Based on {\tool}, we developed a Chrome-based plug-in to facilitate the understanding of the Bash code for developers. 

\item We shared our corpus and scripts in our project homepage\footnote{\url{https://github.com/NTDXYG/BASHEXPLAINER}} to facilitate follow-up studies on the Bash code comment generation problem.
\end{itemize}

\section{Related Work}
\label{sec:relatedwork}

In this section, we first analyze the related studies on generating Bash code from natural language descriptions. Then we analyze the previous studies on automatic code comment generation.
Finally, we emphasize the novelty of our study.

\subsection{Generating Bash Code From Natural Language Description}

Due to the irregular syntax of Bash language, mapping natural language to corresponding Bash commands (i.e., NL2Bash) is a challenging task and has attracted the attention of researchers. 
For example, Lin et al.~\cite{lin2018nl2bash} were the first to address this problem and constructed a corpus of Bash code with natural language descriptions. 
Specifically, they collected the corpus related to Bash code from related repositories (such as developer Q\&A forums, technical tutorials, technical websites, and course materials). After data preprocessing, they collected over 9,000 pairs of data covering over 100 Bash utilities. Finally, they considered three neural machine translation models (i.e., Seq2Seq model~\cite{cho2014learning ,sutskever2014sequence}, CopyNet model~\cite{gu2016incorporating}, and Tellina model~\cite{lin2017program}) and evaluated this task at three different granularities. Kan et al.~\cite{kan2020grid} proposed a new GSAM (Grid Structure Attention Mechanism) mechanism as part of the Seq2Seq model based on the study of Lin et al.~\cite{lin2017program}. Specifically, the GSAM mechanism uses  Bi\text{-}GRU (Bidirectional GRU)~\cite{jain1999recurrent,schuster1997bidirectional} as a feature extraction model, maps these hidden states into a grid structure through a neural network, and uses a convolutional neural network to compute the adjacency features. They further used a copy mechanism~\cite{gu2016incorporating} to alleviate the OOV (out of vocabulary) problem. Compared with the method proposed by Lin et al.~\cite{lin2017program}, this method can improve the performance by 7.0 and 7.3 percentage points in terms of BLEU-1 and BLEU-3 respectively.

\subsection{Automatic Code Comment Generation}

The information retrieval method was one of the earliest kinds of methods since code reuse~\cite{kim2005empirical,kamiya2002ccfinder} is more common in large-scale code repositories.  For corpora with high code reuse, the use of information retrieval methods can achieve better performance. For example, Haiduc et al.~\cite{haiduc2010use} first used VSM (Vector Space Model) and LSI (Latent Semantic Index) to retrieve relevant terms from the corpus to construct code comments. 
Eddy et al.~\cite{eddy2013evaluating} proposed the topic model hPAM to select relevant terms from the corpus to construct code comments. Wong et al.~\cite{wong2013autocomment} used code snippets and corresponding descriptions from the developer Q\&A site Stack Overflow. Specifically, they used the token-based code clone detection tool SIM to detect similar code and used the comments of the detected similar code as the final comments. Later, Wong et al.~\cite{wong2015clocom} proposed the method CloCom. Specifically, they used a token-based code clone detection tool to retrieve similar code from GitHub and used the information from the reviews to generate comments. Liu et al.~\cite{liu2018neural} proposed a nearest neighbor-based method NNGen, where they used a bag-of-words model to convert the code commits in the training set into vectors, and subsequently, selected the $k$ code commits that are most similar to the code commits based on the cosine similarity. The BLEU-4 metric scores of the new code commit with these $k$ code commits are calculated, and the commit message of the code commit with the highest BLEU-4 score is used as the commit message for the new code commit.

To improve the generalization ability of the models, deep learning methods have been used in recent studies. These studies treat the code comment generation task as the neural machine translation (NMT) task~\cite{bahdanau2014neural}. Then the source code is used as the input and the code comment as the output, and deep learning methods are used to train the model. Iyer et al.~\cite{iyer2016summarizing} first proposed CODE-NN, which used LSTM and the attention mechanism to construct encoders and decoders. Hu et al.~\cite{hu2018deep} proposed the method DeepCom, which used the abstract syntax tree to analyze the semantic and structural information of Java code and then proposed the SBT (structure-based traversal) method to traverse the abstract syntax tree and convert the abstract syntax tree into an AST sequence. Ahmad et al.~\cite{ahmad2020transformer} used the Transformer model for code comment generation. Transformer model~\cite{vaswani2017attention} is based on the Multi-headed Self-attention mechanism, which can efficiently capture long-range dependencies. Yang et al.~\cite{yang2021comformer} proposed a Transformer-based method ComFormer, which uses the lexical and syntactic information of code snippets to effectively learn the code semantics. 

Some recent studies proposed hybrid methods (such as the combination of information retrieval methods and deep learning methods) and achieved promising performances. Wei et al.~\cite{wei2020retrieve} proposed the method Re$^2$com, which used the comment of the similar code snippet as an exemplar. This method used the information of exemplar, target code, AST information, and similar codes to help the neural network model to generate code comments. Zhang et al.~\cite{zhang2020retrieval}  proposed the method Rencos, which retrieved the two most similar code snippets from the code repository by considering both syntactic similarity and semantic similarity. Then the target code was encoded and these two codes retrieved similar code snippets and generated comments by fusing them in the decoding phase. Li et al.~\cite{li2021editsum} considered the code 
comment obtained by information retrieval as prototypes, and combined the pattern information in the prototypes with the semantic information of the input code, and then automatically edited the prototypes by the trained neural network to generate code comment. 

\subsection{Novelty of Our Study}

To our best knowledge, previous  studies~\cite{wei2020retrieve,liu2018neural,hu2018deep,yang2021comformer,hu2021automating,li2022setransformer,yang2022ccgir,li2021secnn} mainly focused on popular advanced programming languages (such as Java and Python) and few studies focused on the code comment generation for the script language Bash. 
In this study, we are the first to study the Bash code comment generation problem and propose a novel retrieval-augmented 
comment generation method based on fine-tuned CodeBERT.
Our method employs the two-stage training strategy.
In the first stage, we train the Bash encoder by fine-tuning CodeBERT on our constructed Bash corpus, which can generate higher-quality code vector representation for Bash.
In the second stage, we first retrieval the most similar code based on semantic and lexical similarity. 
Then we generate the code vector representations of the target code and the most similar code  by the Bash encoder. Finally, we use the fusion layer proposed by Yang et al.~\cite{yang2019simple} to fuse these two vectors and finally generate comment based on the decoder.

\section{Our proposed method {\tool}}
\label{sec:method}

\begin{figure*}
	\centering
    \vspace{-2mm}
	\includegraphics[width=0.8\textwidth]{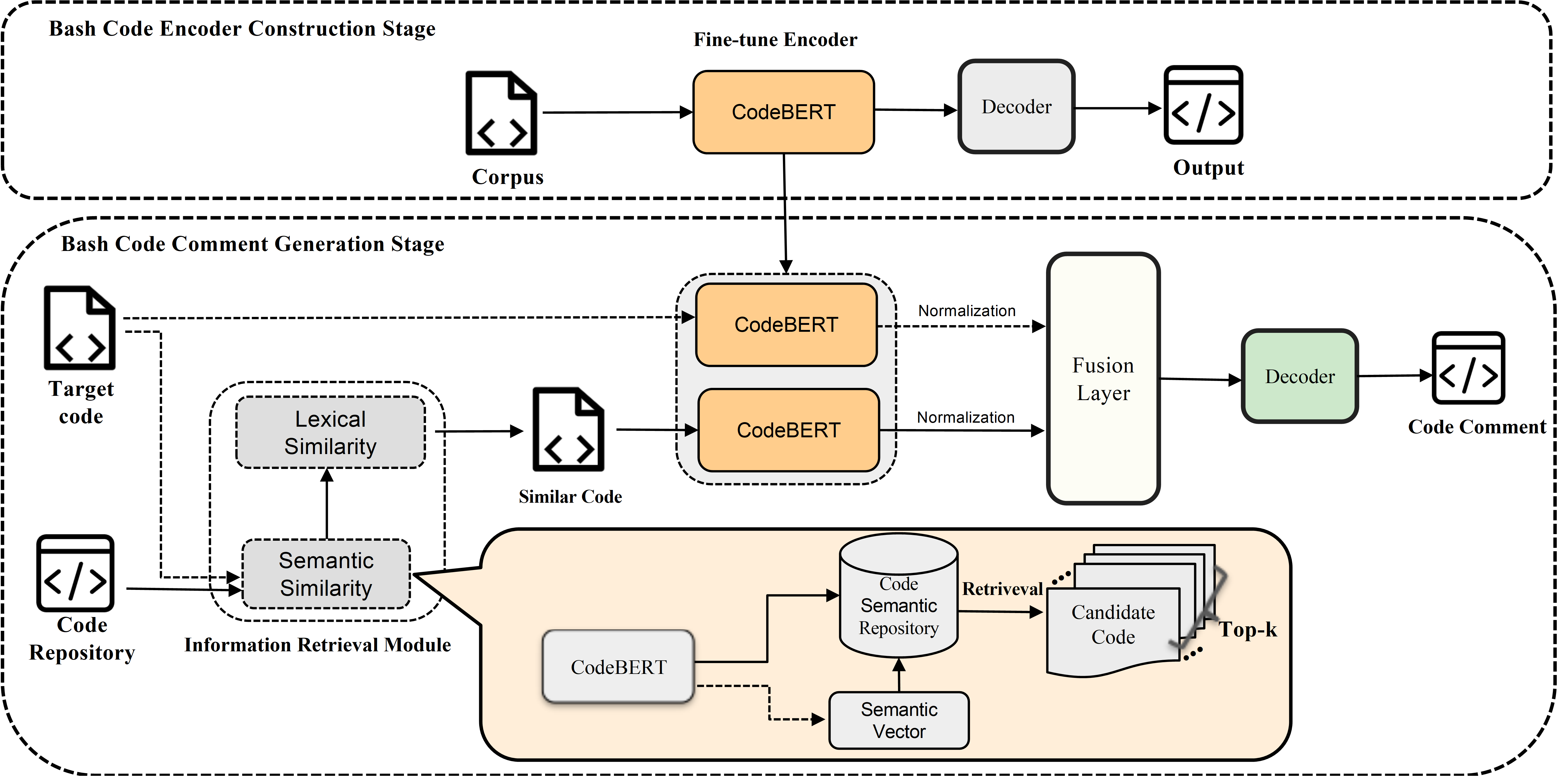}
	\caption{The framework of our proposed method {\tool}}
    \vspace{-2mm}
	\label{fig:Hybrid-ExplainBash}
\end{figure*}

The framework of {\tool} is shown in Figure~\ref{fig:Hybrid-ExplainBash}.
Specifically, {\tool} utilizes two-stage training strategy: {\stageone} and {\stagetwo}.
In the first stage, we want to train the encoder for the Bash code, which can generate higher-quality vector representation  of the Bash code. In this stage, we fine-tune the pre-trained model CodeBERT~\cite{feng2020codebert,zhou2021assessing} on our constructed Bash corpus.
In the second stage, given the target code, we first use the information retrieval module to retrieval the most similar code from the code repository. Then we generate vector representations for these two codes via the Bash code encoder. Later we perform normalization operation and then fuse these two vectors via the fusion layer~\cite{yang2019simple}. Finally, we generate the comment via the decoder 
module.

\subsection{Bash Code Encoder Construction Stage}
CodeBert~\cite{feng2020codebert} is a bi-modal pre-training model for programming languages (PL) and natural languages (NL) based on the encoder in Transformer. 
CodeBert is pre-trained on a large-scale general-purpose corpus by using two tasks: Masked Language Model (MLM) and Replaced Token Detection (RTD). Specifically, the MLM task targets bi-modal data by simultaneously feeding the code with the corresponding comments and randomly selecting positions for masking, then replacing the token with a special token [MASK]. The goal of the MLM task is to predict the original token. The RTD task targets uni-modal data with separate codes and comments and randomly replaces the token, which aims to learn whether the token is the original word via a discriminator.
Specifically, in this stage, we first build a standard Seq2Seq model via CodeBERT as the encoder. Then we fine-tune the CodeBert using our constructed Bash code corpus. In the fine-tuning process, we freeze the model parameters in CodeBERT. Finally, we train the Bash code encoder, which can generate better semantic representational information for the Bash code, and this Bash code encoder can be used in the second stage.

\subsection{Bash Code Comment Generation Stage}

\subsubsection{Information Retrieval Module}

Code reuse~\cite{wei2020retrieve,gharehyazie2017some} is a common practice in software development and similar codes usually have similar code comments. For the target code, we can try to retrieve a similar code from the code repository. However, the syntax of Bash commands with the command options is irregular and it is difficult to represent the Bash code in the abstract syntax tree~\cite{lin2018nl2bash}. 
Therefore, we  construct an information retrieval module by considering both the semantic similarity and the lexical similarity. Here, we first retrieve top-$k$ similar codes from the code repository according to the semantic similarity and then further retrieve the most similar code from these $k$ codes according to the lexical similarity. The rationality of this retrieval order setting is evaluated in Section~\ref{sec:analysis4RQ3}.  

\noindent\textbf{Semantic similarity-based retrieval.}
Considering the promising performance of CodeBERT in natural language search and code document generation tasks~\cite{feng2020codebert}, we use the fine-tuned encoder based on CodeBERT to obtain the semantic vector representations for Bash codes. Then, the semantic similarity is calculated by the Euclidean distance. Later, the top-$k$ candidate codes are retrieved from the code repository based on the semantic similarity. 

The detailed process is illustrated as follows. First the Bash code $X_{tar}$ is input into the fine-tuned encoder based on CodeBERT. Then, the first layer $h^0$ and the last layer $h^n$ of the hidden state are extracted, summed, and averaged to obtain the semantic feature vector $V_{tar}$, where $n$ is the number of hidden layers.

\begin{equation}
V_{tar}=avg(h^0+h^n)
\end{equation}

\modify{Later, we can represent the Bash codes in the code repository }into a vector set ${\{V_i\}}_{i=1}^N$, where $N$ is the number of Bash codes in the code repository. 
Finally, we can compute $semantic\text{-}similarity$ between the target code vector $V_{tar}$ and the historical code vector $V_i$ via the Euclidean distance.

\begin{equation}
semantic\text{-}similarity(V_{tar},V_i)=\sqrt{\sum\limits_{j=1}^{n}(V_{tar}[j]-V_i[j])^2}
\end{equation}

Based on the computed $semantic\text{-}similarity$, top-$k$ candidate Bash codes $\{X\}_{i=1}^k$ can be retrieved in the code repository.

\noindent\textbf{Lexical similarity-based retrieval.} 
We follow the study of Liu et al.~\cite{liu2020retrieval} to calculate the lexical similarity by using text edit distance. First, the set of candidate codes $\{X\}_{i=1}^k$ is traversed, and we treat the traversed code $X_i$ and the target code $X_{tar}$ as the set of tokens. Then we can calculate the $lexical\text{-}similarity$ between them via the text edit distance.

\begin{equation}
lexical\text{-}similarity(X_{tar},X_i)=1-\frac{dis(X_{tar},X_i)}{max(|X_{tar}|,|X_i|)}
\end{equation}
where $dis(X_{tar},X_i)$ is the text edit distance and $|X_{tar}|$ is the length of token sequence of $X_{tar}$. Finally the most similar code $X_{sim}$ is retrieved from the candidate codes based on $lexical\text{-}similarity$.

\subsubsection{Normalization Operation}

After retrieving the similar code $X_{sim}$, we use the fine-tuned encoder based on CodeBERT to obtain the semantic  vector representations $M_{tar}$ and $M_{sim}$ ($M \in \mathbb{R}^{{batch} \times N \times d_{{model}}}$), where the first dimension is the size of the batch, the second dimension is the length of the input sequence, and the third dimension is the vector of embedding representations corresponding to the words from the target code  $X_{tar}$ and the similar code  $X_{sim}$.
Then we want to normalize the third dimension of two representation vectors, which can speed up the convergence of model training and improve the performance of the trained model. Here we follow the previous studies~\cite{henry2020query} and use the $L$2 Normalization method to process the representation vectors, which can result in the normalized vectors $V_{tar}$ and $V_{sim}$.

\begin{equation}
V_{tar}=\frac{M_{tar}}{\left\|M_{tar}\right\|}= \frac{M_{tar}} {\sqrt{\sum\limits_{i=1}^{t}{M_{tar}}_i^2}}
\end{equation}

\begin{equation}
V_{sim}=\frac{M_{sim}}{\left\|M_{sim}\right\|}= \frac{M_{sim}} {\sqrt{\sum\limits_{i=1}^{t}{M_{sim}}_i^2}}
\end{equation}
where $t$ is the dimension of the vector $M$.

\subsubsection{Fusion Layer} 

The fusion layer is used to efficiently fuse the two vectors $V_{tar}$ and $V_{sim}$. In this study, we utilize the fusion layer designed by Yang et al.~\cite{yang2019simple} due to its simplicity and effectiveness, which include three splicing methods.

\begin{equation}
G_1=F_1([V_{tar};V_{sim}])
\end{equation}

\begin{equation}
G_2=F_2([V_{tar};V_{tar}-V_{sim}])
\end{equation}

\begin{equation}
G_3=F_3([V_{tar};V_{tar}\circ V_{sim}])
\end{equation}

Where $F_1$, $F_2$, and $F_3$ are three single-layer feedforward neural networks with mutually independent parameters. $[;]$ means the splicing of two vectors. $-$ denotes the subtraction of two vectors, which can highlight the difference between the two vectors. $\circ$ means the dot product of two vectors, which can highlight the similarity between the two vectors. Then we directly splice the three obtained vectors, input them into another feedforward neural network and then compute the output $G$ of the fusion layer as follows:

\begin{equation}
G=F([G_1;G_2;G_3])
\end{equation}

\subsubsection{Decoder Module}

In the decoder module, we use Transformer's decoder pair as the framework's decoder. The autoregressive mechanism is used in the decoder, which can predict the next possible word based on the previous content.  The specific implementation is that only the left part of the current word is known when decoding. Therefore, the masking mechanism is used to shield the influence of the right part of the current word and maintain the characteristics of autoregression. Each layer of the decoder performs additional cross-attention calculations on the final hidden layer of the encoder and is connected by cross attention. Each decoder layer performs an attention operation on the final hidden state of the encoder output, which can make the output of the generated model closer to the original output.

Specifically, the decoder uses the sentence sequence to predict the next word. That is the previous output is input to the Mask Multi-Head Attention layer. The function of Mask Multi-Head Attention is to block the following words to prevent information leakage. The next steps are almost the same as those in the encoder. The output vector is first passed to the Residual Connection and Layer Normalization layer, then is entered into the Feed-forward layer and is performed residual connection and layer normalization. The above steps are repeated $N$ times, where $N$ represents the number of layers of the decoder. Finally, the output of the decoder $h_t$ is sent to the fully connected neural network, which is then passed to the softmax layer to predict the probability of the next token.

\begin{equation}
P\left(y_{t+1} \mid y_{1}, \cdots, y_{t}\right)=\operatorname{softmax}\left(h_{t} W+b\right)
\end{equation}
where $y$ denotes the predicted token. We train our model parameters $\theta$ by \modify{the  loss function $\mathcal{L}$ based on cross entropy:}

\begin{equation}
\mathcal{L}=-\sum _ { i = 1 } ^ {|y|} {\log P_\theta(y_i|y \textless i,x)}
\end{equation}

Previous studies~\cite{hu2018summarizing,liu2019neural,hu2018deep} showed that using neural networks' maximum probability distribution to generate text often leads to low-quality results. 
Recently, most studies~\cite{wiseman2016sequence,vijayakumar2016diverse,freitag2017beam,cao2021automated} used beam search to achieve better performance on text generation tasks. 
Therefore, in our proposed method {\tool}, we also use the beam search to generate comments for Bash codes.
Specifically, the beam search is a compromise between the greedy strategy and the exhaustive strategy. It retains top-$k$ high-probability words at each step of the prediction as to the input for the next time step, where $k$ denotes the beam size. The larger the value of $k$, the greater the possibility of achieving better performance, but at the cost of more computational cost.
\section{Experimental Setup}
\label{sec:setup}

\subsection{Experimental Subject}
\label{sec:corpus analysis}

To construct our corpus, we first consider the corpus shared by NL2Bash~\cite{lin2018nl2bash}.
NL2Bash~\cite{lin2018nl2bash} was the first study to map natural language descriptions to Bash commands in Linux and shared a high-quality corpus. Then the follow-up studies~\cite{kan2020grid,trizna2021shell} used this corpus as their experimental subject. To better evaluate the performance of our proposed method, we also augmented our corpus with the corpus shared by the NLC2CMD competition. After augmentation, we  removed the duplicated samples from them. Finally, we constructed a larger-scale corpus, which contains a total of 10,592 samples and the composition of each sample in the corpus is $<$Bash code, code comment$>$.

Table~\ref{tab:code length} shows the length statistics of the Bash code and the corresponding comments in the corpus. 
In these tables, we can find that the length of most Bash codes in the corpus is less than 20 and mainly around 8, and the length of the corresponding comments is mainly around 11. In our empirical study, \modify{ we  use a random sampling method to split the dataset into the training set, the validation set, and the test set in the ratio of 80\%: 10\%: 10\% by following previous studies~\cite{yang2021fine,liu2020retrieval,liguori2021shellcode_ia32}}.

\begin{table}[htbp]
\label{tab:test}
	\centering
\caption{Length statistics of samples in the corpus}   
\begin{tabular}{cccccc}    
\toprule 
\multicolumn{6}{c}{Code length statistics}\\    
\midrule  Average  & Mode & Median &$<$16&$<$32&$<$48 \\
8.528 & 4 & 7 &0.908 & 0.997 &0.999 \\ 
\midrule
\multicolumn{6}{c}{Code comment length statistics}\\
\midrule  Average  & Mode & Median &$<$16&$<$32&$<$48 \\
11.874 & 10 & 11 & 0.803 & 0.995 &0.999 \\ 
\bottomrule   
\end{tabular}  
\label{tab:code length}
\end{table}

\subsection{Performance Measures}

To compare the performance between {\tool} and baselines, we consider three performance measures (i.e., BLEU~\cite{papineni2002bleu}, METEOR~\cite{banerjee2005meteor}, and ROUGE-L~\cite{rouge2004package}). These performance measures have been widely used in previous studies for neural machine translation and automatic code comment generation~\cite{hu2018deep,hu2020deep,zhang2020retrieval,wei2020retrieve,yang2022dualsc,liu2022sotitle,yang2021fine}. 
We illustrate the details of these performance measures as follows.

\noindent\textbf{BLEU.}  BLEU (Bilingual Evaluation Understudy)~\cite{papineni2002bleu} is one of the first proposed measures for machine translation quality evaluation. 
BLEU is a precision-based similarity measure for analyzing the degree of simultaneous occurrence of $n$-grams between the candidate text and the reference text. 
The common metrics are BLEU-1, BLEU-2, BLEU-3, and BLEU-4, where $n$-gram refers to the number of consecutive words of $n$.

\noindent\textbf{METEOR.} METEOR (Metric for Evaluation of Translation with Explicit Ordering)~\cite{banerjee2005meteor} is based on a single-precision weighted summed average and single-word recall. METEOR is designed to address the shortcomings in the BLEU measure. 




\noindent\textbf{ROUGE-L.}  ROUGE-L (Recall-Oriented Understudy for Gisting Evaluation)~\cite{rouge2004package} is a measure based on Recall. It is used to calculate the length of the longest common subsequence between the candidate text and the reference text. The longer the length of this sequence, the higher the score of ROUGE-L. 



To avoid result differences due to different implementations, we use the implementations provided by the nlg-eval library\footnote{https://github.com/Maluuba/nlg-eval} for three performance measures, which can mitigate the threat to the internal validity.

\subsection{Baselines}

Since there are no previous studies on Bash code comment generation, to show the competitiveness of our proposed method {\tool}, we compare our proposed method with state-of-the-art baselines from previous studies on code comment generation. 
Specifically, our chosen baselines can cover three different groups. 
The first group contains information retrieval methods and we consider the following four baselines.

\begin{itemize}
    \item \textbf{LSI}~\cite{haiduc2010supporting} retrieves similar codes by calculating the distance between the text and the words in the corpus.
    \item \textbf{VSM}~\cite{haiduc2010use} uses the feature vector of codes for retrieval and uses cosine similarity to retrieve similar codes from the training set. 
    \item \textbf{BM25}~\cite{zhang2020retrieval} is a bag-of-words retrieval function for estimating the correlation between documents and a given search query.
    \item \textbf{NNGen}~\cite{liu2018neural} generates commit messages based on nearest neighbors, it sorts the code in terms of both cosine similarity and BLEU value. 
\end{itemize}

The second group contains deep learning methods and we consider the following four baselines.

\begin{itemize}
    \item \textbf{CopyNet}~\cite{gu2016incorporating} is the model from the NL2Bash study~\cite{lin2018nl2bash} with the best performance, which invokes the copy mechanism into the encoder-decoder structure. 
    
    \item \textbf{Transformer}~\cite{ahmad2020transformer} is an encoder-decoder framework based on Multi-head-Self-attention Mechanism and Position Encoding. Transformer have been widely used in different NLP understanding and generation tasks with promising results. 
    
    \item \textbf{CODE-NN}~\cite{iyer2016summarizing} is the first deep learning model to be used for the comment generation task, which uses LSTM and attention mechanisms to generate code comments.
    
    \item \textbf{CodeBERT}~\cite{feng2020codebert} uses the pre-trained CodeBERT model as an encoder to build the encoder-decoder models. 
\end{itemize}

The third group contains hybrid methods and we consider the following two baselines.

\begin{itemize}
    \item \textbf{Hybrid-Deepcom}~\cite{hu2020deep} combines the lexical and structural information of Java methods and considers the syntactic information of the code by performing structural traversal of the AST of the code.
    \item \textbf{Rencos}~\cite{zhang2020retrieval} is a retrieval-based neural source code summary generation method that uses retrieved similar codes to augment the neural model. 

\end{itemize}

For the baselines  (i.e., CopyNet~\cite{gu2016incorporating}, Transformer~\cite{ahmad2020transformer}, and CodeBERT~\cite{feng2020codebert}) that do not share code, we implemented them  according to the method description and the results of our implementations are very close to the results reported in the original studies. For the remaining baselines, we directly utilized the scripts shared by original studies. To ensure a fair comparison between {\tool} and these baselines, we tuned and optimized the hyper-parameters in these methods.

\subsection{Experimental Settings}

Our proposed method and some baselines were implemented based on the PyTorch framework. We use the packages Faiss\footnote{\url{https://github.com/facebookresearch/faiss}}, Textdistance\footnote{\url{https://github.com/life4/textdistance}} and Transformers\footnote{\url{https://github.com/huggingface/transformers}} to implement our proposed method. \modify{Specifically, we use the Levenshtein edit distance~\cite{levenshtein1966binary} from the Textdistance package to calculate lexical similarity, and use the Adamw optimizer for 30 epochs with a learning rate of 2e-4 for the model training. The values of the hyperparameters are optimized according to our best practices
and are shown in Table~\ref{tab:Hyper-parameters}.}

We run all the experiments on a computer with an Intel(R) Xeon(R) Silver 4210 CPU and a GeForce RTX3090 GPU with 24 GB memory. The running OS platform is Windows OS.

\begin{table}[htbp]
 \caption{Hyper-parameters of {\tool} and their values}
 \begin{center}
\begin{tabular}{ccc}
\toprule
        \textbf{Category}               & \textbf{Hyper-parameter} & \textbf{Value} \\ \midrule
\multirow{5}{*}{Deep Learning Part} & decoder\_layers       & 6     \\
                        & hidden\_size        & 768     \\
                       & max\_input\_length        & 64     \\
                       &max\_output\_length         & 32  \\
                       &beam\_search\_size        & 10  \\
                      \midrule
\multirow{2}{*}{Information Retrieval Part} & top-$k$         & 8  \\
                       & CodeBERT\_hidden\_size         & 768    \\

  \bottomrule
\end{tabular}
 \end{center}
 \label{tab:Hyper-parameters}
\end{table}

\subsection{Tool implementation}

\begin{figure}[htbp]
	\centering
  \vspace{-1mm}
	\includegraphics[width=0.4\textwidth]{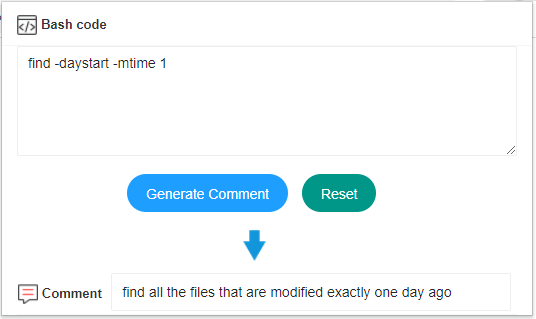}
	\caption{The screenshot of our developed Chrome browser plug-in}
  \vspace{-1mm}
	\label{fig:tool}
\end{figure}

To facilitate the developers to understand Bash codes, we implemented a tool based on {\tool} and integrated it into the Chrome browser. The screenshot of our developed tool is shown in Figure~\ref{fig:tool}. After calling out the plug-in,
developers can copy the Bash code into the input box and click the ``Generate Comment" button to quickly generate the corresponding comment.

\section{Result Analysis}
\label{sec:result}

\subsection{Result Analysis for RQ1}

\noindent\textbf{RQ1: {\RQone}}

In this RQ, we want to evaluate the performance of {\tool} in an automated way. Since Bash code comment generation has not been investigated in previous studies, we consider 10 state-of-the-art baselines from similar problems (such as code comment generation). These baselines can be divided into three groups: information retrieval methods, deep learning methods, and hybrid methods. Since measuring the similarity between human-generated code comments and model-generated comments is a challenging task,
we used three performance measures~\cite{papineni2002bleu,banerjee2005meteor,rouge2004package}, which have been widely used in previous studies of neural machine translation and code comment generation.

\begin{table*}[]
 \caption{Comparison results between our proposed method {\tool} and baselines}
 \begin{center}
  \vspace{-1mm}
 \setlength{\tabcolsep}{1mm}{
  \resizebox{1.0\textwidth}{!}{
\begin{tabular}{c|c|cccccc}
  \toprule
\textbf{Method Type} & \textbf{Method Name} & \textbf{$\mathit{BLEU}$-$1$ (\%)} & \textbf{$\mathit{BLEU}$-$2$ (\%)} & \textbf{$\mathit{BLEU}$-$3$ (\%)} & \textbf{$\mathit{BLEU}$-$4$ (\%)} & \textbf{$\mathit{METEOR}$ (\%)} & \textbf{$\mathit{ROUGE}$-$\mathit{L}$ (\%)}\\
  \midrule
\multirow{4}{*}{Information Retrieval}
& LSI & 30.18 & 18.07 & 12.48 & 9.40 & 18.30 & 28.82  \\
& VSM & 36.16 & 24.47 & 18.62 & 15.25 & 22.04 & 34.58   \\
& BM25 & 42.08 & 30.41 & 23.58 & 19.24 & 26.35 & 38.49 \\
& NNGen & 50.62 & 38.75 & 32.11  & 27.85 & 27.69 & 45.88 \\
\midrule
\multirow{4}{*}{Deep Learning} & 
CopyNet & 38.11  & 27.06 & 20.67 & 16.43 & 22.06  & 40.18 \\
& Transformer & 46.39 & 33.37 & 25.42 & 19.97 & 25.22 & 44.01 \\
& CODE-NN & 49.60 & 37.18 & 29.53 & 24.17 & 26.85 & 47.21\\
& CodeBERT & 48.65 & 37.02 & 29.84 & 24.83 & 27.16 & 47.36 \\
\midrule
\multirow{2}{*}{Hybrid Method} &
Hybrid-DeepCom & 47.78 & 35.45 & 27.91 &22.75 & 26.27 & 45.36 \\
& Rencos & 46.27 & 35.11 & 28.66 & 24.39 & 25.82 & 45.06\\
\midrule
\textbf{Our Method}&{\tool} &\textbf{52.61} &\textbf{41.54}  & \textbf{34.92} & \textbf{30.44} & \textbf{29.01} & \textbf{49.19}\\
  \bottomrule
\end{tabular}}
 }
 \end{center}
 \label{tab:RQ1result}
\end{table*}


Table~\ref{tab:RQ1result} shows the automatic evaluation results of our proposed method {\tool} and our considered baselines. 
In this table, we can find that our proposed method {\tool} can achieve the best performance in terms of all performance measures. 

Firstly, we compare {\tool} with the information retrieval baselines. We find that the performance of the information retrieval methods is not high except for the NNGen method. These results show that using simple information retrieval methods is not suitable for the Bash comment generation task. The NNGen method uses cosine similarity as well as BLEU values to retrieve similar code, and this method can retrieve code that is more similar to the target code, so the NNGen method can achieve a significant improvement in BLEU values. However, compared with the NNGen method, {\tool} can achieve higher performance in all performance measures.  Specifically, in terms of BLEU-1/2/3/4 measure, {\tool} can improve the performance by at least
3.93\%, 7.20\%, 8.75\%, and 9.29\%. In terms of METEOR and ROUGE-L measures, {\tool} can also improve 4.77\% and 7.21\%. \yc{The possible reason is that it is difficult to guarantee that the retrieved similar code and the target code can be perfectly matched by only using the information retrieval method for our studied Bash code comment generation problem.}

Secondly, we compare {\tool} with the deep learning baselines.
 We select CopyNet~\cite{gu2016incorporating} as the baseline, since CopyNet can achieve the best performance in the NL2Bash study. Here we want to investigate whether this method is suitable for Bash code comment generation and the comparison results show the unsatisfactory performance of the CopyNet method. Then, we find that the CODE-NN method  and the CodeBERT method  can achieve the best performance among all deep learning baselines. 
 However, our proposed method {\tool} can improve the performance by at least 6.07\%, 11.73\%, 17.02\%, and 22.59\% in terms of the BLEU-1/2/3/4 measures, while in terms of the METEOR and ROUGE-L measures, {\tool} can also improve the performance by at least 6.81\% and 3.86\%. 
 Compared with the information retrieval method NNGen, the CODE-NN method  and the CodeBERT method  perform worse than the NNGen method in most performance measures. This indicates that for Bash code comment generation, only using deep learning methods cannot help to achieve promising performance. Since most deep learning models require a large amount of training data, the size of our constructed corpus for Bash code is still small when compared to the corpus for Java or Python~\cite{hu2018deep,wan2018improving}. We conjecture this is one possible reason why the deep learning method does not perform well in our study. In addition, according to the statistical information of the corpus shown in Section~\ref{sec:corpus analysis}, the average length of Bash code is 8.528. Due to the short length of most Bash code snippets, it is challenging for deep learning  methods to learn semantic information in short texts, which may also lead to poor performance of deep learning methods.

Finally, we compare our proposed method with the hybrid baselines. Final comparison results show that our proposed method {\tool} can improve the performance by at least 10.10\%, 17.18\%, 21.84\%, and 24.80\% in terms of the BLEU-1/2/3/4 measures, while the method {\tool} can also improve the performance by at least 10.43\% and 8.44\% in terms of the METEOR and ROUGE-L measures. 

\subsection{Result Analysis for RQ2}

\noindent\textbf{RQ2: {\RQtwo}}

In this RQ, we want to investigate whether using the two-stage training strategy is help for improve the performance.
Therefore, we design a control method \textbf{w/o two-stage training} for comparison.  In this control method, the hyper-parameters' values of the Bash code encoder, the fusion layer, and decoder are optimized through a single-stage training strategy.

\begin{table*}[]
 \caption{Comparison results for {\tool} with or without the two-stage training strategy}
 \begin{center}
  \vspace{-1mm}
 \setlength{\tabcolsep}{1mm}{
  \resizebox{1.0\textwidth}{!}{
\begin{tabular}{c|cccccc}
  \toprule
\textbf{Setting} & \textbf \textbf{$\mathit{BLEU}$-$1$ (\%)} & \textbf{$\mathit{BLEU}$-$2$ (\%)} & \textbf{$\mathit{BLEU}$-$3$ (\%)} & \textbf{$\mathit{BLEU}$-$4$ (\%)} & \textbf{$\mathit{METEOR}$ (\%)} & \textbf{$\mathit{ROUGE}$-$\mathit{L}$ (\%)}\\
  \midrule
 w/o two-stage training & 46.35 & 35.77 & 29.26 & 24.75 & 26.83 & 48.06 \\
{\tool} &\textbf{52.61} &\textbf{41.54}  & \textbf{34.92} & \textbf{30.44} & \textbf{29.01} & \textbf{49.19}\\
  \bottomrule
\end{tabular}}
 }
 \end{center}
 \label{tab:RQ2Result}
\end{table*}

Table~\ref{tab:RQ2Result} shows the final comparison results. 
In this table, we can find that  {\tool} can significantly improve the performance in terms of all performance measures by considering the two-stage training strategy. Specifically, using the two-stage training strategy can improve the performance 13.5\%, 16.1\%, 19.3\%, and 22.9\% in terms of the BLEU-1/2/3/4 measures, respectively.
In terms of the METEOR and ROUGE-L measures,  using the two-stage training strategy can improve the performance  8.1\% and 2.3\% respectively. Therefore, construction Bash code encoder by  fine-tuning CodeBERT can help to significantly improve the performance of {\tool} and the two-stage training strategy  is necessary for our proposed method {\tool}.

\subsection{Result Analysis for RQ3}
\label{sec:analysis4RQ3}

\noindent\textbf{RQ3: {\RQthree}}

In this RQ, we aim to design a set of ablation experiments to verify the component setting rationality of our proposed method {\tool}. 
To show the component setting rationality of our proposed method {\tool}, we design a set of control methods.
In particular,
to analyze the rationality of the information retrieval module in {\tool},
we replace our information retrieval module with the best information retrieval baseline NNGen and keep the remaining modules unchanged (denoted as \textbf{with NNGen}). 
To show the rationality of the similarity retrieval order in our information retrieval module,
we switched the retrieval order in the similarity retrieval module (denoted as \textbf{with Reverse Retrieval}). 
To show the setting rationality of the normalization operation,
we remove the normalization operation from {\tool} (denoted as \textbf{ w/o Normalization}).  
To show the setting rationality of the fusion layer,
we replaced our considered fusion layer that contains three splicing methods with the simple fusion method that directly splices two semantic vectors (denoted as \textbf{with Simple Fusion}).
Finally, to show the competitiveness of using the encoder-decoder structure in {\tool}, we remove the encoder-decoder structure and only use the information retrieval module (denoted as \textbf{ w/o NMT}).

\begin{table*}[]
 \caption{Ablation study results for our proposed method {\tool}}
 \begin{center}
  \vspace{-1mm}
 \setlength{\tabcolsep}{1mm}{
  \resizebox{1.0\textwidth}{!}{
\begin{tabular}{c|cccccc}
  \toprule
\textbf{Setting} & \textbf \textbf{$\mathit{BLEU}$-$1$ (\%)} & \textbf{$\mathit{BLEU}$-$2$ (\%)} & \textbf{$\mathit{BLEU}$-$3$ (\%)} & \textbf{$\mathit{BLEU}$-$4$ (\%)} & \textbf{$\mathit{METEOR}$ (\%)} & \textbf{$\mathit{ROUGE}$-$\mathit{L}$ (\%)}\\
  \midrule
 with NNGen & 50.74 & 39.22 & 32.37 & 27.75 & 28.26 & 48.41 \\
 with Reverse Retrieve & 51.85 & 40.86 & 34.35 & 30.02 & 28.72 & 48.41 \\
w/o Normalization &51.13 & 40.38 & 34.05 & 29.69 & 28.78 & 49.10 \\
with  Simple Fusion &51.51 & 40.44 & 33.84 & 29.01 & 28.02 & 48.21 \\
w/o NMT &51.81 &40.52  & 33.96 & 29.62 & 28.45 & 47.76\\
{\tool} &\textbf{52.61} &\textbf{41.54}  & \textbf{34.92} & \textbf{30.44} & \textbf{29.01} & \textbf{49.19}\\
  \bottomrule
\end{tabular}}
 }
 \end{center}
 \label{tab:RQ3Result}
\end{table*}



 Table~\ref{tab:RQ3Result} shows the results of the ablation experiments. After comparing these control methods, our proposed method can achieve the best performance.  Specifically,
 compared with \textbf{with NNGen} and \textbf{with Reverse Retrieval}, \yc{we can find that our proposed information retrieval module is reasonable and the order setting of first retrieval by semantic similarity and second retrieval by lexical similarity can help to retrieve similar codes more effectively than other methods.}
 Compared with \textbf{ w/o Normalization}, \yc{we can verify the rationality of the normalization operation (i.e., normalizing the representation vector can improve the performance of {\tool}).} 
 Compared with \textbf{with Simple Fusion}, \yc{the result shows that the fusion layer can effectively fuse the representational information of the similar code and the target code, which can finally improve the quality of the generated code comments. This also shows  the fusion layer designed by Yang et al.~\cite{yang2019simple} is still effective for Bash code comment generation task.}
 Compared with \textbf{ w/o NMT}, \yc{we can find that when only using the information retrieval method, it is difficult to ensure that the reused code comment can perfectly match the target code in terms of code semantics.}

\section{Discussion}
\label{sec:discussiond}

\subsection{Comment Quality Comparison by Human Study}
\label{sec:humanstudy}

Using automatic measures can only assess the lexical gap between the generated comments and the human-written comments. However, it does not truly reflect the semantic gap between the generated code comments and the human-written comments~\cite{iyer2016summarizing,wei2020retrieve,zhang2020retrieval,hu2021automating}. Therefore, we conducted a human study to measure the quality of comments generated by NNGen~\cite{liu2018neural}, CodeBERT~\cite{feng2020codebert}, Hybrid-DeepCom~\cite{hu2020deep} and  {\tool}, since these selected three baselines have the best performance in each group of baselines. In our human study, we followed the methodology used by previous studies on source code comment generation~\cite{wei2020retrieve,li2021editsum}. 

In our human study, we measure the quality of the comments in three different perspectives:

\begin{itemize}
    \item \textbf{Similarity}. This perspective concerns the similarity between the generated comment and the human-written comment.
    \item \textbf{Naturalness}. This perspective concerns the grammaticality and fluency of the generated comment.
    \item \textbf{Informativeness}. This perspective concerns the amount of content carried over from the Bash code to the generated comment, which ignores the fluency of the comment.
\end{itemize}

To measure the quality of the generated comments, we first hired five master students, who have extensive experience in using Bash for Linux system development and maintenance. Then we randomly selected 100 Bash codes from the test set, which include human-written comments and the comments generated by four methods. \modify{We provide the human-written comment since the hired students can score generated comments by referring this ground truth.} Each student was asked to rate the four comments for each Bash code based on similarity, naturalness, and informativeness on a scale from 0 to 4, with higher scores indicating that the code comments can better meet the requirements. A sample of our used questionnaire can be found in our project homepage. During the review process, students can search the Internet for relevant information and unfamiliar concepts. To guarantee a fair comparison, students do not know which comment is generated by which method, \modify{ and the order of questionnaires is different for different students}. To guarantee the label quality, we need each student to review only 50 Bash codes in half a day.

\begin{table}[htbp]
\label{tab:test}
\caption{The comparison results (standard deviation in parentheses) of our human study}   
\begin{tabular}{cccc}    
\toprule 
 \textbf{Method} & \textbf{Informativeness}& \textbf{Naturalness}& \textbf{Similarity}\\
 \midrule
NNGen & 1.78 (1.32) & \textbf{3.61} (0.89) &1.44 (1.65) \\   
CodeBERT & 2.11 (1.22) & 3.10 (1.18) &1.66 (1.63)\\ 
Hybrid-DeepCom & 2.37 (1.54) &3.32 (1.02) &2.01 (1.25)\\ 
{\tool} & \textbf{2.74} (1.35) & 3.54 (0.76) &\textbf{2.41} (1.30)\\ 
\bottomrule   
\end{tabular}  
\label{tab:discussion1}
\end{table}

Table~\ref{tab:discussion1} shows the the human study results between {\tool} and three representative baselines. In this table, we can find {\tool} can outperform the other three baselines by at least 15.6\%  in terms of Informativeness perspective and 19.9\%  in terms of Similarity perspective, respectively. It is worth noting that the performance in terms of Naturalness perspective is only 0.07 points lower than the NNGen method  because the code comments generated by the  NNGen method are retrieved from the code repository. The code comments in the code repository are all written by developers, so they mainly conform to the natural language syntax and have higher naturalness. The performance of {\tool} is slightly lower than that of the  NNGen method, which  shows that {\tool}  can also generate comments with higher naturalness. 
Similar to the results of previous studies~\cite{wei2020retrieve,hu2021automating}, we used standard deviation and $p$-values to further evaluate the experimental comparison results. The differences in the standard deviations of the four methods are small, which means there are relatively few differences in scoring among students.  All $p$-values were less than 0.05, which means the comparison result difference exists statistically significant.
Finally, we used Fleiss Kappa~\cite{fleiss1971measuring} to measure the consistency of the scoring results between these five students. The final Fleiss Kappa value is 0.723, which indicate that there exists consistency in the scoring results of these students.

\subsection{ Qualitative Analysis} 

Table~\ref{tab:discussion2} shows the code comments generated by four methods for three Bash codes in the test set. 
In our empirical study, we can find that the information retrieval method NNGen can perform well in automatic evaluation. However, directly reusing the comment from the corpus may not be suitable for Bash code.
For example, due to the language flexibility of Bash code, similar Bash codes may have different semantics. For the last two Bash codes, the retrieved Bash code by NNGen has a large semantic difference from the target code.  
Moreover, we find the CodeBERT method can generate higher-quality comments than the Hybrid-DeepCom method.
However, the generated comment still has a certain distance from the ground truth. 
For example, for the second Bash code, the comment generated by CodeBERT misses the phrase ``under current directory" while the comment generated by {\tool} can match the ground-truth comment. The reason is that {\tool} can effectively fuse information from the similar code and the target code.


\begin{table}[]
\caption{Generated comments of different methods for three Bash codes}
 \resizebox{0.5\textwidth}{!}{
\begin{tabular}{l|l}
\toprule
\textbf{ID}   &\textbf{Example}                  \\
\midrule
\textbf{1} & \begin{tabular}[c]{@{}l@{}}

\textbf{Bash Code:} find . -type f -name *.php.\\  

\textbf{Ground Truth:} find all php files under current directory \\
\textbf{Hybrid-DeepCom:} search current directory tree for regular file\\ whose name end php\\
\textbf{CodeBERT:} find all php file under current directory\\
\textbf{NNGen:} list all files/directories under current directory using \\comma (,) as the delimiter for different fields in the output \\
\textbf{{\tool}:} find all php files under current directory\\
\end{tabular}\\
\midrule
\textbf{2} & \begin{tabular}[c]{@{}l@{}}

\textbf{Bash Code:} find . -type l -! -exec test -e {} $\backslash$;  -print.\\  

\textbf{Ground Truth:} find all broken symlinks under current directory \\
\textbf{Hybrid-DeepCom:} convert all broken symlink under current directory\\
\textbf{CodeBERT:} find broken symbol link\\
\textbf{NNGen:} search the `images' directory tree for regular files \\
\textbf{{\tool}:} find all broken symlinks under current directory\\
\end{tabular}\\
 \midrule
 \textbf{3} & \begin{tabular}[c]{@{}l@{}}

\textbf{Bash Code:} unset array[`shuf -i 0-3 -n1'].\\  
\textbf{Ground Truth:} unsets random one from first four array members. \\
\textbf{Hybrid-DeepCom:} unset random one in first \\
\textbf{CodeBERT:} unset random one from first four array member\\
\textbf{NNGen:}search directory /users/david/desktop/ recursively for\\ regular files with extensions .txt, .mpg, .jpg\\
\textbf{{\tool}:} unset random one from first four array members\\
\end{tabular}\\
  \bottomrule

\end{tabular}
}
\label{tab:discussion2}
\end{table}

\section{Threats to validity}
\label{sec:threats}

In this section, we analyze potential threats to the validity of empirical study.

\noindent\textbf{Threats to internal validity.} 
To avoid faults in the implementations, we check our implementation carefully and use mature libraries. For example, we use the Faiss library for information retrieval and the PyTorch framework for implementing {\tool} and baselines. The second threat is the considered baselines. Since the Bash code comment generation problem has not been investigated in previous studies, we mainly choose the state-of-the-art baselines from the source code comment generation domain and aim to cover different types of methods, such as information retrieval methods, deep learning methods, and hybrid methods.

\noindent\textbf{Threats to external validity.} 
The main external threat is the representativeness of the choice corpus. To alleviate this threat, we first consider the corpus provided by NL2Bash\cite{lin2018nl2bash} and the competition data of the NLC2CMD Challenge. Then we merged these two corpora to construct a larger-scale corpus and removed the duplicated Bash codes in these two corpora.

\noindent\textbf{Threats to construct validity.} 
The structural threats concern the performance measures used to evaluate the performance of {\tool} and baselines. To mitigate these threats, we chose three performance measures BLEU~\cite{papineni2002bleu}, METEOR~\cite{banerjee2005meteor}, and ROUGE-L~\cite{rouge2004package}. These measures have been widely used in previous studies on neural machine translation and source code comment generation. Moreover, we also conducted a human study to analyze the generated comment quality in the manual way.

\noindent\textbf{Threats to conclusion validity.} 
\modify{
Due to the high computational cost of deep learning and the sufficient number of samples in our constructed corpus, we only split  the dataset once. This setting is consistent with the previous  studies on source code summarization~\cite{wei2020retrieve,liu2020retrieval,zhang2020retrieval}.
To alleviate the threat to the conclusion validity, we also randomly split our datasets three times with different random seeds. Due to the limitation of paper length, we show the detailed comparison result on  our project homepage and the comparison results also confirm the effectiveness of our proposed method.}

\section{Conclusion and Future Work}
\label{sec:conclusion}

In this study, we are the first to investigate the problem of automatic Bash code comment generation and then propose the retrieval-augmented neural source code comment generation method {\tool} based on the fine-tuned CodeBERT. The results of the automated evaluation and human study show that our proposed method {\tool} can outperform the
state-of-the-art baselines from the previous studies of automatic code comment generation on our constructed Bash code corpus. Moreover, we also verify the component setting rationality of {\tool} by designing a set of ablation experiments. 

In the future, we first want to augment our corpus by gathering more Bash code from GitHub and Stack Overflow, which can help to cover more types of Bash commands. We second want to improving the quality of generated comments by designing more effective fusion methods.

\section*{Acknowledgment}
The authors would like to thank the anonymous reviewers for their insightful comments and suggestions. 
Chi Yu and
Guang Yang have contributed equally to this work and they
are co-first authors.
This work is supported in part by the National Natural Science Foundation of
China (Grant no. 61872263).

\bibliographystyle{IEEEtran}
\bibliography{mylib}

\begin{thebibliography}{10}
\providecommand{\url}[1]{#1}
\csname url@samestyle\endcsname
\providecommand{\newblock}{\relax}
\providecommand{\bibinfo}[2]{#2}
\providecommand{\BIBentrySTDinterwordspacing}{\spaceskip=0pt\relax}
\providecommand{\BIBentryALTinterwordstretchfactor}{4}
\providecommand{\BIBentryALTinterwordspacing}{\spaceskip=\fontdimen2\font plus
\BIBentryALTinterwordstretchfactor\fontdimen3\font minus
  \fontdimen4\font\relax}
\providecommand{\BIBforeignlanguage}[2]{{%
\expandafter\ifx\csname l@#1\endcsname\relax
\typeout{** WARNING: IEEEtran.bst: No hyphenation pattern has been}%
\typeout{** loaded for the language `#1'. Using the pattern for}%
\typeout{** the default language instead.}%
\else
\language=\csname l@#1\endcsname
\fi
#2}}
\providecommand{\BIBdecl}{\relax}
\BIBdecl

\bibitem{lin2018nl2bash}
X.~V. Lin, C.~Wang, L.~Zettlemoyer, and M.~D. Ernst, ``Nl2bash: A corpus and
  semantic parser for natural language interface to the linux operating
  system,'' in \emph{Proceedings of the Eleventh International Conference on
  Language Resources and Evaluation (LREC 2018)}, 2018.

\bibitem{sridhara2010towards}
G.~Sridhara, E.~Hill, D.~Muppaneni, L.~Pollock, and K.~Vijay-Shanker, ``Towards
  automatically generating summary comments for java methods,'' in
  \emph{Proceedings of the IEEE/ACM international conference on Automated
  software engineering}, 2010, pp. 43--52.

\bibitem{wei2020retrieve}
B.~Wei, Y.~Li, G.~Li, X.~Xia, and Z.~Jin, ``Retrieve and refine: exemplar-based
  neural comment generation,'' in \emph{2020 35th IEEE/ACM International
  Conference on Automated Software Engineering (ASE)}.\hskip 1em plus 0.5em
  minus 0.4em\relax IEEE, 2020, pp. 349--360.

\bibitem{xia2017measuring}
X.~Xia, L.~Bao, D.~Lo, Z.~Xing, A.~E. Hassan, and S.~Li, ``Measuring program
  comprehension: A large-scale field study with professionals,'' \emph{IEEE
  Transactions on Software Engineering}, vol.~44, no.~10, pp. 951--976, 2017.

\bibitem{liu2018neural}
Z.~Liu, X.~Xia, A.~E. Hassan, D.~Lo, Z.~Xing, and X.~Wang,
  ``Neural-machine-translation-based commit message generation: how far are
  we?'' in \emph{Proceedings of the 33rd ACM/IEEE International Conference on
  Automated Software Engineering}, 2018, pp. 373--384.

\bibitem{hu2018deep}
X.~Hu, G.~Li, X.~Xia, D.~Lo, and Z.~Jin, ``Deep code comment generation,'' in
  \emph{2018 IEEE/ACM 26th International Conference on Program Comprehension
  (ICPC)}.\hskip 1em plus 0.5em minus 0.4em\relax IEEE, 2018, pp. 200--20\,010.

\bibitem{yang2021comformer}
G.~Yang, X.~Chen, J.~Cao, S.~Xu, Z.~Cui, C.~Yu, and K.~Liu, ``Comformer: Code
  comment generation via transformer and fusion method-based hybrid code
  representation,'' in \emph{2021 8th International Conference on Dependable
  Systems and Their Applications (DSA)}.\hskip 1em plus 0.5em minus 0.4em\relax
  IEEE, 2021, pp. 30--41.

\bibitem{hu2021automating}
X.~Hu, Z.~Gao, X.~Xia, D.~Lo, and X.~Yang, ``Automating user notice generation
  for smart contract functions,'' in \emph{2021 36th IEEE/ACM International
  Conference on Automated Software Engineering (ASE)}.\hskip 1em plus 0.5em
  minus 0.4em\relax IEEE, 2021, pp. 5--17.

\bibitem{li2022setransformer}
Z.~Li, Y.~Wu, B.~Peng, X.~Chen, Z.~Sun, Y.~Liu, and D.~Paul, ``Setransformer: A
  transformer-based code semantic parser for code comment generation,''
  \emph{IEEE Transactions on Reliability}, 2022.

\bibitem{yang2022ccgir}
G.~Yang, K.~Liu, X.~Chen, Y.~Zhou, C.~Yu, and H.~Lin, ``Ccgir: Information
  retrieval-based code comment generation method for smart contracts,''
  \emph{Knowledge-Based Systems}, vol. 237, p. 107858, 2022.

\bibitem{li2021secnn}
Z.~Li, Y.~Wu, B.~Peng, X.~Chen, Z.~Sun, Y.~Liu, and D.~Yu, ``Secnn: A semantic
  cnn parser for code comment generation,'' \emph{Journal of Systems and
  Software}, vol. 181, p. 111036, 2021.

\bibitem{feng2020codebert}
Z.~Feng, D.~Guo, D.~Tang, N.~Duan, X.~Feng, M.~Gong, L.~Shou, B.~Qin, T.~Liu,
  D.~Jiang \emph{et~al.}, ``Codebert: A pre-trained model for programming and
  natural languages,'' in \emph{Findings of the Association for Computational
  Linguistics: EMNLP 2020}, 2020, pp. 1536--1547.

\bibitem{gu2016incorporating}
J.~Gu, Z.~Lu, H.~Li, and V.~O. Li, ``Incorporating copying mechanism in
  sequence-to-sequence learning,'' in \emph{Proceedings of the 54th Annual
  Meeting of the Association for Computational Linguistics (Volume 1: Long
  Papers)}, 2016, pp. 1631--1640.

\bibitem{vaswani2017attention}
A.~Vaswani, N.~Shazeer, N.~Parmar, J.~Uszkoreit, L.~Jones, A.~N. Gomez,
  {\L}.~Kaiser, and I.~Polosukhin, ``Attention is all you need,''
  \emph{Advances in neural information processing systems}, vol.~30, 2017.

\bibitem{iyer2016summarizing}
S.~Iyer, I.~Konstas, A.~Cheung, and L.~Zettlemoyer, ``Summarizing source code
  using a neural attention model,'' in \emph{Proceedings of the 54th Annual
  Meeting of the Association for Computational Linguistics (Volume 1: Long
  Papers)}, 2016, pp. 2073--2083.

\bibitem{haiduc2010use}
S.~Haiduc, J.~Aponte, L.~Moreno, and A.~Marcus, ``On the use of automated text
  summarization techniques for summarizing source code,'' in \emph{2010 17th
  Working Conference on Reverse Engineering}.\hskip 1em plus 0.5em minus
  0.4em\relax IEEE, 2010, pp. 35--44.

\bibitem{haiduc2010supporting}
S.~Haiduc, J.~Aponte, and A.~Marcus, ``Supporting program comprehension with
  source code summarization,'' in \emph{2010 acm/ieee 32nd international
  conference on software engineering}, vol.~2.\hskip 1em plus 0.5em minus
  0.4em\relax IEEE, 2010, pp. 223--226.

\bibitem{hu2020deep}
X.~Hu, G.~Li, X.~Xia, D.~Lo, and Z.~Jin, ``Deep code comment generation with
  hybrid lexical and syntactical information,'' \emph{Empirical Software
  Engineering}, vol.~25, no.~3, pp. 2179--2217, 2020.

\bibitem{zhang2020retrieval}
J.~Zhang, X.~Wang, H.~Zhang, H.~Sun, and X.~Liu, ``Retrieval-based neural
  source code summarization,'' in \emph{2020 IEEE/ACM 42nd International
  Conference on Software Engineering (ICSE)}.\hskip 1em plus 0.5em minus
  0.4em\relax IEEE, 2020, pp. 1385--1397.

\bibitem{cho2014learning}
K.~Cho, B.~Van~Merri{\"e}nboer, C.~Gulcehre, D.~Bahdanau, F.~Bougares,
  H.~Schwenk, and Y.~Bengio, ``Learning phrase representations using rnn
  encoder-decoder for statistical machine translation,'' \emph{arXiv preprint
  arXiv:1406.1078}, 2014.

\bibitem{sutskever2014sequence}
I.~Sutskever, O.~Vinyals, and Q.~V. Le, ``Sequence to sequence learning with
  neural networks,'' \emph{Advances in neural information processing systems},
  vol.~27, 2014.

\bibitem{lin2017program}
X.~V. Lin, C.~Wang, D.~Pang, K.~Vu, and M.~D. Ernst, ``Program synthesis from
  natural language using recurrent neural networks,'' \emph{University of
  Washington Department of Computer Science and Engineering, Seattle, WA, USA,
  Tech. Rep. UW-CSE-17-03-01}, 2017.

\bibitem{kan2020grid}
J.-W. Kan, W.-C. Chien, and S.-D. Wang, ``Grid structure attention for natural
  language interface to bash commands,'' in \emph{2020 International Computer
  Symposium (ICS)}.\hskip 1em plus 0.5em minus 0.4em\relax IEEE, 2020, pp.
  67--72.

\bibitem{jain1999recurrent}
L.~C. Jain and L.~R. Medsker, \emph{Recurrent neural networks: design and
  applications}.\hskip 1em plus 0.5em minus 0.4em\relax CRC Press, Inc., 1999.

\bibitem{schuster1997bidirectional}
M.~Schuster and K.~K. Paliwal, ``Bidirectional recurrent neural networks,''
  \emph{IEEE transactions on Signal Processing}, vol.~45, no.~11, pp.
  2673--2681, 1997.

\bibitem{kim2005empirical}
M.~Kim, V.~Sazawal, D.~Notkin, and G.~Murphy, ``An empirical study of code
  clone genealogies,'' in \emph{Proceedings of the 10th European software
  engineering conference held jointly with 13th ACM SIGSOFT international
  symposium on Foundations of software engineering}, 2005, pp. 187--196.

\bibitem{kamiya2002ccfinder}
T.~Kamiya, S.~Kusumoto, and K.~Inoue, ``Ccfinder: A multilinguistic token-based
  code clone detection system for large scale source code,'' \emph{IEEE
  transactions on software engineering}, vol.~28, no.~7, pp. 654--670, 2002.

\bibitem{eddy2013evaluating}
B.~P. Eddy, J.~A. Robinson, N.~A. Kraft, and J.~C. Carver, ``Evaluating source
  code summarization techniques: Replication and expansion,'' in \emph{2013
  21st International Conference on Program Comprehension (ICPC)}.\hskip 1em
  plus 0.5em minus 0.4em\relax IEEE, 2013, pp. 13--22.

\bibitem{wong2013autocomment}
E.~Wong, J.~Yang, and L.~Tan, ``Autocomment: Mining question and answer sites
  for automatic comment generation,'' in \emph{2013 28th IEEE/ACM International
  Conference on Automated Software Engineering (ASE)}.\hskip 1em plus 0.5em
  minus 0.4em\relax IEEE, 2013, pp. 562--567.

\bibitem{wong2015clocom}
E.~Wong, T.~Liu, and L.~Tan, ``Clocom: Mining existing source code for
  automatic comment generation,'' in \emph{2015 IEEE 22nd International
  Conference on Software Analysis, Evolution, and Reengineering (SANER)}.\hskip
  1em plus 0.5em minus 0.4em\relax IEEE, 2015, pp. 380--389.

\bibitem{bahdanau2014neural}
D.~Bahdanau, K.~Cho, and Y.~Bengio, ``Neural machine translation by jointly
  learning to align and translate,'' \emph{arXiv preprint arXiv:1409.0473},
  2014.

\bibitem{ahmad2020transformer}
W.~Ahmad, S.~Chakraborty, B.~Ray, and K.-W. Chang, ``A transformer-based
  approach for source code summarization,'' in \emph{Proceedings of the 58th
  Annual Meeting of the Association for Computational Linguistics}, 2020, pp.
  4998--5007.

\bibitem{li2021editsum}
J.~Li, Y.~Li, G.~Li, X.~Hu, X.~Xia, and Z.~Jin, ``Editsum: A retrieve-and-edit
  framework for source code summarization,'' in \emph{2021 36th IEEE/ACM
  International Conference on Automated Software Engineering (ASE)}.\hskip 1em
  plus 0.5em minus 0.4em\relax IEEE, 2021, pp. 155--166.

\bibitem{yang2019simple}
R.~Yang, J.~Zhang, X.~Gao, F.~Ji, and H.~Chen, ``Simple and effective text
  matching with richer alignment features,'' in \emph{Proceedings of the 57th
  Annual Meeting of the Association for Computational Linguistics}, 2019, pp.
  4699--4709.

\bibitem{zhou2021assessing}
X.~Zhou, D.~Han, and D.~Lo, ``Assessing generalizability of codebert,'' in
  \emph{2021 IEEE International Conference on Software Maintenance and
  Evolution (ICSME)}.\hskip 1em plus 0.5em minus 0.4em\relax IEEE, 2021, pp.
  425--436.

\bibitem{gharehyazie2017some}
M.~Gharehyazie, B.~Ray, and V.~Filkov, ``Some from here, some from there:
  Cross-project code reuse in github,'' in \emph{2017 IEEE/ACM 14th
  International Conference on Mining Software Repositories (MSR)}.\hskip 1em
  plus 0.5em minus 0.4em\relax IEEE, 2017, pp. 291--301.

\bibitem{liu2020retrieval}
S.~Liu, Y.~Chen, X.~Xie, J.~K. Siow, and Y.~Liu, ``Retrieval-augmented
  generation for code summarization via hybrid gnn,'' in \emph{International
  Conference on Learning Representations}, 2020.

\bibitem{henry2020query}
A.~Henry, P.~R. Dachapally, S.~S. Pawar, and Y.~Chen, ``Query-key normalization
  for transformers,'' in \emph{Findings of the Association for Computational
  Linguistics: EMNLP 2020}, 2020, pp. 4246--4253.

\bibitem{hu2018summarizing}
X.~Hu, G.~Li, X.~Xia, D.~Lo, S.~Lu, and Z.~Jin, ``Summarizing source code with
  transferred api knowledge,'' in \emph{Proceedings of the 27th International
  Joint Conference on Artificial Intelligence}, 2018, pp. 2269--2275.

\bibitem{liu2019neural}
B.~Liu, T.~Wang, X.~Zhang, Q.~Fan, G.~Yin, and J.~Deng, ``A neural-network
  based code summarization approach by using source code and its call
  dependencies,'' in \emph{Proceedings of the 11th Asia-Pacific Symposium on
  Internetware}, 2019, pp. 1--10.

\bibitem{wiseman2016sequence}
S.~Wiseman and A.~M. Rush, ``Sequence-to-sequence learning as beam-search
  optimization,'' in \emph{Proceedings of the 2016 Conference on Empirical
  Methods in Natural Language Processing}, 2016, pp. 1296--1306.

\bibitem{vijayakumar2016diverse}
A.~K. Vijayakumar, M.~Cogswell, R.~R. Selvaraju, Q.~Sun, S.~Lee, D.~Crandall,
  and D.~Batra, ``Diverse beam search: Decoding diverse solutions from neural
  sequence models,'' \emph{arXiv preprint arXiv:1610.02424}, 2016.

\bibitem{freitag2017beam}
M.~Freitag and Y.~Al-Onaizan, ``Beam search strategies for neural machine
  translation,'' in \emph{Proceedings of the First Workshop on Neural Machine
  Translation}, 2017, pp. 56--60.

\bibitem{cao2021automated}
K.~Cao, C.~Chen, S.~Baltes, C.~Treude, and X.~Chen, ``Automated query
  reformulation for efficient search based on query logs from stack overflow,''
  in \emph{2021 IEEE/ACM 43rd International Conference on Software Engineering
  (ICSE)}.\hskip 1em plus 0.5em minus 0.4em\relax IEEE, 2021, pp. 1273--1285.

\bibitem{trizna2021shell}
D.~Trizna, ``Shell language processing: Unix command parsing for machine
  learning,'' \emph{arXiv preprint arXiv:2107.02438}, 2021.

\bibitem{yang2021fine}
G.~Yang, Y.~Zhou, X.~Chen, and C.~Yu, ``Fine-grained pseudo-code generation
  method via code feature extraction and transformer,'' in \emph{2021 28th
  Asia-Pacific Software Engineering Conference (APSEC)}.\hskip 1em plus 0.5em
  minus 0.4em\relax IEEE, 2021, pp. 213--222.

\bibitem{liguori2021shellcode_ia32}
P.~Liguori, E.~Al-Hossami, D.~Cotroneo, R.~Natella, B.~Cukic, and S.~Shaikh,
  ``Shellcode\_ia32: A dataset for automatic shellcode generation,'' in
  \emph{Proceedings of the 1st Workshop on Natural Language Processing for
  Programming (NLP4Prog 2021)}, 2021, pp. 58--64.

\bibitem{papineni2002bleu}
K.~Papineni, S.~Roukos, T.~Ward, and W.-J. Zhu, ``Bleu: a method for automatic
  evaluation of machine translation,'' in \emph{Proceedings of the 40th annual
  meeting of the Association for Computational Linguistics}, 2002, pp.
  311--318.

\bibitem{banerjee2005meteor}
S.~Banerjee and A.~Lavie, ``Meteor: An automatic metric for mt evaluation with
  improved correlation with human judgments,'' in \emph{Proceedings of the acl
  workshop on intrinsic and extrinsic evaluation measures for machine
  translation and/or summarization}, 2005, pp. 65--72.

\bibitem{rouge2004package}
L.~C. ROUGE, ``A package for automatic evaluation of summaries,'' in
  \emph{Proceedings of Workshop on Text Summarization of ACL, Spain}, 2004.

\bibitem{yang2022dualsc}
G.~Yang, X.~Chen, Y.~Zhou, and C.~Yu, ``Dualsc: Automatic generation and
  summarization of shellcode via transformer and dual learning,'' in
  \emph{Proceedings of The 29th IEEE International Conference on Software
  Analysis, Evolution and Reengineering (SANER 2022)}, 2022.

\bibitem{liu2022sotitle}
K.~Liu, G.~Yang, X.~Chen, and C.~Yu, ``Sotitle: A transformer-based post title
  generation approach for stack overflow,'' in \emph{Proceedings of The 29th
  IEEE International Conference on Software Analysis, Evolution and
  Reengineering (SANER 2022)}, 2022.

\bibitem{levenshtein1966binary}
V.~I. Levenshtein \emph{et~al.}, ``Binary codes capable of correcting
  deletions, insertions, and reversals,'' in \emph{Soviet physics doklady},
  vol.~10, no.~8.\hskip 1em plus 0.5em minus 0.4em\relax Soviet Union, 1966,
  pp. 707--710.

\bibitem{wan2018improving}
Y.~Wan, Z.~Zhao, M.~Yang, G.~Xu, H.~Ying, J.~Wu, and P.~S. Yu, ``Improving
  automatic source code summarization via deep reinforcement learning,'' in
  \emph{Proceedings of the 33rd ACM/IEEE International Conference on Automated
  Software Engineering}, 2018, pp. 397--407.

\bibitem{fleiss1971measuring}
J.~L. Fleiss, ``Measuring nominal scale agreement among many raters.''
  \emph{Psychological bulletin}, vol.~76, no.~5, p. 378, 1971.

\end{thebibliography}


\end{document}